%% file: WCNC_datft_v6.tex
\documentclass[conference]{IEEEtran}
\IEEEoverridecommandlockouts 

\usepackage{verbatim} 
\usepackage{graphicx, caption}
\usepackage{amssymb}
\usepackage{amsmath}
\usepackage{amsthm}
\usepackage{mathtools}
\usepackage{cite}
\usepackage{stfloats}
\usepackage{epstopdf}
\usepackage{psfrag}
\usepackage[mathscr]{euscript}
\usepackage{acronym}  
\usepackage{booktabs} 
\usepackage{algorithmic}
\usepackage[linesnumbered, ruled,vlined]{algorithm2e}
\usepackage[table]{xcolor}
\usepackage{subcaption}

\SetKwInput{KwInput}{Input}
\SetKwInput{KwOutput}{Output}
\SetKwComment{Comment}{/* }{ */}

\usepackage{color}
\usepackage{dsfont}
\usepackage{bbm}

\input{SupportDocuments/acronym}

\input{SupportDocuments/defmetric}

\allowdisplaybreaks 

\begin{document}
	\newcommand{\paperTitle}{Title}
\vspace{-0.0cm}


\title{\vspace{-0.0cm} A Bargaining Game for Personalized, Energy Efficient Split Learning over Wireless Networks}

\author{ \vspace{-0.05 cm}	
	
\IEEEauthorblockN{
	Minsu~Kim,  Alexander~DeRieux, and Walid~Saad
}\\[-0.5em]
\vspace{-2.0mm}
\IEEEauthorblockA{
Wireless@VT, Bradley Department of Electrical and Computer Engineering, Virginia Tech, Arlington, VA, USA. 
\\Emails: \{msukim, acd1797, walids\}@vt.edu.
	\vspace{-1.0 cm}	
}
\thanks{This work was supported by the U.S. National Science Foundation under Grant CNS-2114267.
\vspace{-1.0mm}
}
}

\maketitle 

%

%

\acresetall
\begin{abstract}

Split learning (SL) is an emergent distributed learning framework which can mitigate the computation and wireless communication overhead of federated learning. It splits a machine learning model into a device-side model and a server-side model at a cut layer. Devices only train their allocated model and transmit the activations of the cut layer to the server. However, SL can lead to data leakage as the server can reconstruct the input data using the correlation between the input and intermediate activations. Although allocating more layers to a device-side model can reduce the possibility of data leakage, this will lead to more energy consumption for resource-constrained devices and more training time for the server. Moreover, non-iid datasets across devices will reduce the convergence rate leading to increased training time. In this paper, a new personalized SL framework is proposed. For this framework, a novel approach for choosing the cut layer that can optimize the tradeoff between the energy consumption for computation and wireless transmission, training time, and data privacy is developed. In the considered framework, each device personalizes its device-side model to mitigate non-iid datasets while sharing the same server-side model for generalization. To balance the energy consumption for computation and wireless transmission, training time, and data privacy, a multiplayer bargaining problem is formulated to find the optimal cut layer between devices and the server. To solve the problem, the Kalai-Smorodinsky bargaining solution (KSBS) is obtained using the bisection method with the feasibility test. Simulation results show that the proposed personalized SL framework with the cut layer from the KSBS can achieve the optimal sum utilities by balancing the energy consumption, training time, and data privacy, and it is also robust to non-iid datasets.

\end{abstract}

\section{Introduction}
\Ac{FL} is a promising solution for distributed inference as it enables multiple devices and a server to train a shared model without revealing private data \cite{HB:16}. Since each device trains a whole model and transmits it to the server iteratively, significant wireless communication and computation overhead can exist on devices. To mitigate this challenge, \ac{SL} was proposed in \cite{Si:19}, In \ac{SL} the model is split into two separate portions, which are a device-side model and a server-side model, at the cut layer. The devices and the server communicate over a wireless channel.  A device only needs to train its allocated model and transmit the activations of the cut layer to the server. Then, the server with more computing resources trains the remaining model based on the received information. However, the server can still reconstruct the private data of the devices from the received activations due to the high correlation between the activations and the input when the allocated device-side model is too shallow \cite{Privacy1D, Privacyleakage}. Although one can reduce the possibility of data leakage by increasing the device-side model, the training will become computationally intensive for resource-constrained devices. In addition, this will increase the training time as the server should wait until devices finish processing their models. Moreover, non-iid datasets across devices will increase the training time by reducing the convergence rate. Thus, it is important to find the optimal cut layer by balancing the energy consumption related to computation and wireless transmission, training time, and data privacy and to develop an algorithm for robust performance over non-iid datasets. 

Several prior works \cite{SplitFed, Privacy1D, Privacyleakage, Mi:21} studied the problems of data privacy and non-iid datasets in \ac{SL} scenarios over communication networks. In \cite{SplitFed}, the authors proposed SplitFed in which device-side training was parallelized and differential privacy was incorporated to improve data privacy. The work in \cite{Privacy1D} demonstrated that data leakage can happen when training convolutional neural networks in \ac{SL}.  In \cite{Privacyleakage}, the authors proposed a novel \ac{SL} algorithm to enhance data privacy by minimizing the distance correlation between the intermediate activations and the input data. Meanwhile, in \cite{Mi:21}, 
the authors studied the use of \ac{SL} at inference stage over wireless networks and the impact of non-iid datasets on its performance.

However, these works  \cite{SplitFed, Privacy1D, Privacyleakage, Mi:21} did not consider the impact of the cut layer on energy consumption, training time, and data privacy. Only few works such as\cite{Han:21} and \cite{Wu:22} considered the optimal cut layer in terms of training latency. The work in \cite{Han:21} developed a local-loss-based training for \ac{SL} and derived the optimal cut layer to minimize the training latency. In \cite{Wu:22}, cluster-based parallel \ac{SL} was proposed along with a resource management algorithm to minimize its training time by optimizing the cut layer selection. To the best of our knowledge, there are no prior works on \ac{SL} that jointly consider energy consumption for computation and communication, training time, and data privacy to obtain the optimal cut layer for devices and the server. 

The main contribution of this paper is a novel personalized \ac{SL} framework that can handle heterogeneous datasets and that is equipped with a new approach to find the optimal cut layer between devices and the server\footnote{The source code is publicly available on https://github.com/news-vt.}. In our personalized \ac{SL} model, the learning model is divided into two separate portions: a device-side model and a server-side model. Each device personalizes its own device-side model while sharing the same server-side model. At the beginning of the learning, each device performs forward propagation on its allocated model in parallel and transmits the activations of the cut layer to the server. Then, the server completes the forward propagation with each device's activations and performs back propagation on its model separately, in parallel. The server transmits the gradients of its last layer to the corresponding devices so that they can finish back propagation. Subsequently, the server performs FedAvg on its updated models to generate a new server-side model. We then formulate utility functions for the devices and the server by capturing energy consumption of computation and communication, training time, and data privacy. In particular, devices can reduce energy consumption by choosing a shallow cut layer. However, this can result in data leakage due to the high correlation between the cut layer's activations and the input data. Meanwhile, the server may want to choose the shallow cut layer so that it can leverage its computing capability to minimize the training time. To capture this conflict over the cut layer between devices and the server, we formulate a multiplayer bargaining problem whose goal is to maximize the utilities of devices and the server. To solve the problem, we obtain the \ac{KSBS} using the bisection method with the feasibility test. Simulation results show that personalized \ac{SL} with the optimal cut layer from the \ac{KSBS} can achieve robust performance over non-iid datasets with fast convergence while achieving the best sum utilities by balancing the energy consumption, training time, and data privacy.

The rest of this paper is organized as follows. Section II presents the system model. In Section III, we formulate the bargaining problem. Section IV provides simulation results.
Finally, conclusions are drawn in Section V.

\begin{figure}
			\includegraphics[width=1.0\columnwidth]{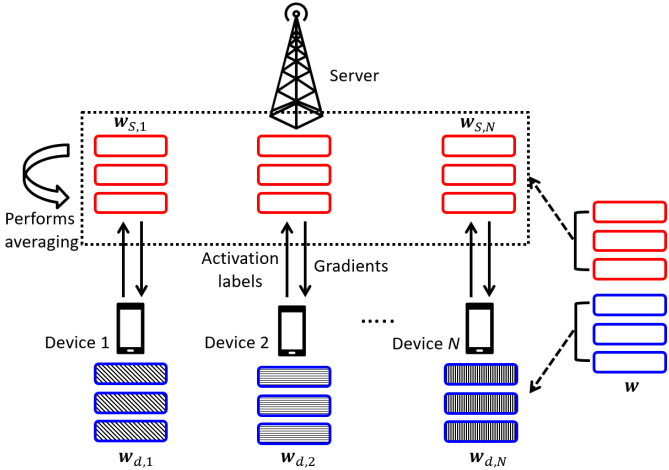}
 			\captionsetup {singlelinecheck = false}

		\caption{An illustration of the personalized \ac{SL} system over wireless networks.}
		\label{fig:System_model}

\vspace{-0.5 cm}	
\end{figure}

\section{System Model} \label{sec:system model}
We consider a personalized \ac{SL} system, in which one server and a set of devices $\deviceset$ with $|\deviceset| = \Whole$ (e.g. mobile or \ac{IoT} devices) collaboratively train a \ac{ML} model to execute a certain data analysis task. All devices have their personalized layers while sharing the same subsequent layers with the server as shown in Fig. \ref{fig:System_model}. The server generates an \ac{ML} model $\weights$ for an image classification task. Let $|\weights|$ be the number of model parameters in the generated model. For device $k$, we define $\dweights$ as the device-side model, $\forall k \in \deviceset$ and $\sweights$ as the server-side model. We use $\fraction$  such that $0\leq \fraction \leq 1$ to allocate $|\dweights| = \fraction |\weights|$, $\forall k \in \deviceset$, model parameters to a device-side model and $|\sweights| = (1-\fraction) |\weights|$ model parameters to the server-side model. Note that all device-side models share the same architecture while they are personalized to each device. The main goal of the personalized \ac{SL} system is to solve the following problem:
\begin{align}
	\min_{\alldweights, \sweights} \frac{1}{\Wholedata} \sum_{k \in \deviceset} \sum_{l \in \dataset} \loss(\dweights, \sweights, \boldsymbol{x}_{kl}, y_{kl}), \label{SL problem}
\end{align} 
where $\alldweights = (\weights_{d,1}, \dots, \weights_{d, \Whole})$,  $\dataset$ is the input dataset of device $k$ with $|\dataset| = \datasize$, $\Wholedata = \sum_{k \in \deviceset} \datasize$ is the total number of data samples across devices, and $\loss(\cdot, \cdot, \cdot, \cdot)$ is a loss function for a given sample. We assume that all devices use the same loss function. $\boldsymbol{x}_{kl}$ is an input vector $l$ of device $k$, and $y_{kl}$ is the corresponding output with $l = \{1, \dots, \datasize\}.$ Without loss of generality, we consider unbalanced and non-iid dataset $\dataset$ across devices.

\subsection{Proposed Personalized \ac{SL} algorithm}
We now describe the proposed algorithm to solve problem \eqref{SL problem}. The server uses FedAvg \cite{HB:16} to train $\sweights$ while each device updates its personalized layers using a gradient based algorithm. For a given $\fraction$, each device $k\in \deviceset$ receives its device-side model $\dweights$ from the server and initializes it. The server also generates $\boldsymbol{w}_{S, k}, \forall k \in \deviceset$. Motivated by \cite{SplitFed} and \cite{Ar:19}, we assume that each device $k \in \deviceset$ performs forward propagation in parallel on $\dweights$ at each local step using mini-batch $\minibatch$. Then, device $k \in \deviceset$ transmits the intermediate outputs, i.e., activations, $\activation$ and the corresponding labels $\dlabels \in \minibatch$ to the server. Based on the received information, the server can finish forward propagation and perform back propagation on $\boldsymbol{w}_{S, k}(t)$. Subsequently, it transmits the gradients of its last layer to the corresponding device. Then, device $k$ can update $\dweights(t)$ using the received gradients. After $\SGDrun$ local steps, the server perform FedAvg on $\boldsymbol{w}_{S, k}(t), \forall k \in \deviceset$, to generate $\sweights(t+1) = \sum_{k \in \deviceset} \frac{\datasize}{\sum_{k \in \deviceset} \datasize} \sdweights(t)$. Then, at the next global round $t+1$, the server sets $\sdweights(t+1) = \sweights(t+1)$, $\forall k \in \deviceset.$ We summarize the aforementioned algorithm in Algorithm 1.

\begin{algorithm}[t!]
	\scriptsize
	\caption{Proposed Personalized \ac{SL} Algorithm 
	}\label{alg:1}
	{
		\While{global round \(t \neq R\)}{
			\If{t = 0}{
				Initialize $\dweights(0)$ and $\sdweights(0)$ $\forall k\in\deviceset$;
			}
			{
				\For{device $k \in \deviceset$}{
					\While{local step $i \neq \SGDrun$}{
						\Comment{Forward Propagation}
						Device $k$ samples mini-batch $\minibatch$
						$\activation \gets forward(\dweights(t), \xi_k)$\;
						Device $k$ transmits \(\activation\) and label $\dlabels$ to server\;
						$\hat{y}_{k} \gets forward(\sdweights(t), \activation)$\;
						\BlankLine
						\BlankLine
						\Comment{Backward Propagation}
						\(\ell_{k}(t) \gets loss(\dlabels, \hat{y}_{k})\)\;
						Server computes $\nabla \ell_{k}(\sdweights(t))$\;
						\(\sdweights(t)  \gets \sdweights(t) - \eta \nabla \ell_{k}(\sdweights(t)) \)\;
						Server transmits gradient of its last layer $\mathrm{d}\activation(t)$ to device $k$ \;
						Using $\mathrm{d} \activation(t)$, device $k$ updates
						\( \dweights(t) \gets \dweights(t) - \eta \nabla \ell_k(\dweights(t))  \)\
					}
				}
				\Comment{FedAvg}
				\(\sweights(t+1) \gets \sum_{k \in \deviceset} \frac{\datasize}{\sum_{k \in \deviceset} \datasize} \sdweights(t) \)\;
				Set $\sdweights(t+1) = \sweights(t+1), \ \forall k \in \deviceset$ \;
			}
		}
	}
\end{algorithm}

\subsection{Wireless Transmission and Computing Model}

\subsubsection{Wireless transmission model}
After device $k$ finishes forward propagation on $\dweights$, it transmits activations $\activation$ and the corresponding labels $\dlabels$ to the server using \ac{OFDMA}. Then, the achievable rate of device $k$ can be given by 
\begin{align}
	\achievable = \bandwidth \log_2 \left(
	1 + \frac{\txpower \channel}{\Noise \bandwidth}
	\right),
\end{align}
where $\bandwidth$ is the bandwidth allocated to device $k$, $\channel$ is the channel gain between device $k$ and the server, $\txpower$ is the transmission power, $\Noise$ is the power spectral density of white Gaussian noise. Then, the transmission time to upload $\activation$ and $\dlabels$ will be
\begin{align}
	\ttime = \frac{|\activation| + |\dlabels|}{\achievable}.
\end{align}
Then, the energy consumption to transmit $\activation$ and $\dlabels$ to the server is $\tenergy = \ttime \txpower$. Since the server usually has a high transmission power and large bandwidth for the downlink, we neglect the energy and the time to transmit the gradients of its last layer \cite{Zhaya:21}. 

\subsubsection{Computing model}
Let $\dfreq$ be the CPU frequency of device $k$. Then the energy consumption to train $\dweights$ for one global round using $\dataset$ will be given by \cite{Han:21}
\begin{align}
	\denergy = \cpuparam \fraction \datasize \cpunumber \dfreq^2,
\end{align}
where $\cpuparam$ is the effective capacitance coefficient of CPU \cite{Ngutra:19}, $\cpunumber$ is the number of required CPU cycles to process one data sample. Note that $\denergy$ is a function of $\fraction$ since device $k$ processes $\dweights$, which has $\fraction$$|\weights|$ number of model parameters. The computation time will be
\begin{align}
	\dtime = \frac{\fraction \cpunumber \datasize}{\dfreq}.
\end{align}
Similarly, we can define the energy consumption of the server for one global round $t$ as $\senergy = \sum_{k \in \deviceset} \datasize (1-\fraction) \cpuparam \cpunumberserver \sfreq^2$, where $\cpunumberserver$ is the number of requires CPU cycles to process one data sample for the server and $\sfreq$ is its CPU frequency. Then, the computation time of the server will be $\stime = \max_{k \in \deviceset} \datasize (1-\fraction) \cpunumberserver/\sfreq$. Since the server processes $\sdweights, \forall k \in \deviceset$ in parallel, $\stime$ will be determined by the largest computation time.

\subsection{Utility Functions}
Now, we define the utility functions of each device and the server. 
Since the server usually has a strong computing capability, it may want to set $\fraction$ small so as to reduce the elapsed time during training. For devices, the optimal $\fraction$ should neither be too small because of the possibility of data leakage nor too large because of the energy consumption for training. Specifically, there exists high probability of data leakage when device-side models are shallow. As $\fraction$ decreases, the correlation between the input data and an intermediate layer output, i.e., activations $\activation$, increases. Hence, it is possible to reconstruct input data from activations as shown in \cite{Privacy1D} and \cite{Privacyleakage}. In other words, an honest-but-curious server can do model inversion attack during training to restore private input data \cite{Modelinversion}. However, training a large device-side model would be also infeasible for resource-constrained devices since training a deep neural network consumes significant energy.

To capture this tradeoff between privacy and energy consumption for devices, we define the utility function of each device $k \in \deviceset$ for one global round as follows 
%
%
\begin{align}
	\dutil = \underbrace{\payoff \dfreq}_{(a)} - (\underbrace{\denergy +\SGDrun\tenergy}_{(b)}) +\underbrace{\privacy \log_2 (1+\fraction)}_{(c)}, \label{dutil}
\end{align}
where $(a)$ is the received reward from the server for the allocated computing resources with payoff $\payoff$, $(b)$ is the energy consumption for training $\dweights$ and transmitting the intermediate outputs to the server, and $(c)$ is a function to measure privacy protection with coefficient $\privacy$ to capture the preference of data privacy. Note that as $\fraction$ increases the correlation between input data and the intermediate outputs become decreased \cite{Privacyleakage}.
We then define the utility function of the server for one global round as below
\begin{align}
	\sutil &= B - \bigg[ \underbrace{\sum_{k \in \deviceset} \payoff \dfreq}_{(a)} + 
	\balance  \underbrace{\senergy}_{(b)} 
	\ka
	& \quad + (1- \balance)
	\underbrace{	\left\{
	\stime + \max_{k \in \deviceset} \dtime + \SGDrun\E[\ttime]
	\right\}
	 }_{(c)}
 	\bigg] \label{sutil}
,
\end{align}
where $\budget$ is the available budget of the server, $(a)$ is the amount of payoff for devices, $(b)$ is the energy consumption for training $\sdweights, \forall k \in \deviceset$, $(c)$ is the elapsed time to compute $\sdweights, \forall k $ and the elapsed time to wait for the slowest device to finish computing its model, $\E(\cdot)$ is with respect to $\channel$ and $\balance$ is a parameter to balance the interests between the energy consumption and the training time. We assume that the server can control $\payoff$ so that $\sutil$ and $\dutil, \forall k $ can be larger than zero. 

From the above utility functions, we can see that devices and the server have conflicting interests over $\fraction$. If the server prioritizes minimizing training time, then it will try to set $\fraction$ as a low value so as to leverage its high computing power. However, when $\fraction$ is low, there exists high probability of data leakage for the devices. Hence, they need to reach a certain agreement for $\fraction$ to initiate personalized \ac{SL}. This situation can be modeled as a bargaining game between devices and the server as they can mutually benefit from reaching the optimal $\fraction^*$ while conflict exists on the terms of the agreement \cite{gametheory}. 

In the following section, we obtain the \ac{KSBS} to find the optimal split.

\section{Personalized SL as a Bargaining Game} \label{sec:problem formulation}
We formulate a bargaining game to reach an agreement over $\fraction$. We first define the set of all feasible utility functions as:
\begin{align}
	\utilset \hspace{-0.2mm} = \hspace{-0.2mm} \left\{
	U_{d,1}(\fraction), \dots, U_{d, \Whole}(\fraction), U_S (\fraction) \ \hspace{-0.2mm} | \hspace{-0.2mm} \ \hspace{-0.2mm} 0 \hspace{-0.2mm}\leq \hspace{-0.2mm} \fraction \hspace{-0.2mm} \leq  \hspace{-0.2mm}1
	\right\}.
\end{align}
Let $\disagreementset = \{\phi_{d,1}, \dots, \phi_{d, \Whole}, \phi_S   \}$ be the disagreement point, which is a set of utilities when devices and the server fail to come to an agreement. Then, our bargaining game can be defined as the pair $(\utilset, \disagreementset)$, and the bargaining solution is a function $\barginingsolution$ that maps $(\utilset, \disagreementset)$ to a unique outcome $\barginingsolution(\utilset, \disagreement) \in \utilset$. Our bargaining solution should prioritize a device with important or private-sensitive dataset so that it can achieve a higher utility than devices with less important datasets. Therefore, while there are many bargaining approaches (e.g., Nash bargaining, etc.), we choose the \ac{KSBS} \cite{gametheory}. This is because the monotonicty axiom of the \ac{KSBS} can capture the aforementioned benefit since a device with a stronger privacy preference $\privacy$ will be able to get a larger achievable maximum utility and a larger utility set. Thus, it can have stronger bargaining power than others leading to a better output $\fraction^*$.


It is known that the \ac{KSBS} is the largest element in $\utilset$ that is on the line connecting $\disagreement$ and $\ideal$, where $\ideal$ is the vector of individually maximized utilities. The \ac{KSBS} point is essentially the solution to the following optimization problem:
\begin{align}
	&\max \quad \KSBS \label{KSBS_problem} \\
	&\ \text{s.t.} \quad \disagreement + \KSBS (\ideal - \disagreement )\in \utilset.
\end{align}
For the disagreement point $\disagreement$, we can set $\disagreement = 0$ because the server cannot initiate the learning if devices and the server fail to negotiate on $\fraction$. Then, we can simplify the problem as 
\begin{align}
	&\max \quad \KSBS \label{simplified_KSBS_problem} \\
	&\ \text{s.t.} \quad  \KSBS\ideal \label{constraint} \in \utilset.
\end{align}
Now, the \ac{KSBS} will lie on the line connecting the origin point and $\ideal$. To solve problem \eqref{simplified_KSBS_problem}, we use the bisection method with a feasibility test to tackle constraint \eqref{constraint}. Firstly, we characterize $\ideal = (U_{d,1}^{\text{ideal}}, U_{d,2}^{\text{ideal}}, \dots, U_{d, \Whole}^{\text{ideal}}, U_{S}^{\text{ideal}}  )$. From \eqref{dutil}, it is straightforward to see that $\dutil$ is concave with respect to $\fraction$ as $\frac{\partial^2 \dutil}{\partial \fraction^2} = -\frac{\privacy \log2}{(1+\fraction)^2} < 0$. Hence, we can obtain $U_{d, k}^{ \text{ideal} }$ from the first derivative test as below 
\begin{align}
	\frac{\partial \dutil}{\partial \fraction} = \frac{\privacy}{\log{2} \times (1+\fraction)} - \cpuparam \datasize \cpunumber \dfreq^2 = 0.
\end{align}
Then, the solution of the above equation can be given by 
\begin{align}
	\hat{\fraction}_k = \frac{\privacy}{\log{2} \times \cpuparam \cpunumber \datasize \dfreq^2} - 1 \label{dfraction}.
\end{align}
From \eqref{dfraction}, we can see that the optimal split ratio $\hat{\fraction}_k$ for device $k$ increases as the preference of data protection $\privacy$ increases. For the $\sutil$, its first derivative can be given by
\begin{align}
	\frac{\partial \sutil}{\partial \fraction} \hspace{-0.5mm} = \hspace{-0.5mm} \gamma \hspace{-1.3mm} \sum_{k \in \deviceset} \hspace{-0.5mm} \cpuparam \datasize \cpunumberserver \sfreq^2 \hspace{-0.5mm} +  \hspace{-0.5mm} (1 \hspace{-0.5mm}- \hspace{-0.5mm} \gamma) \hspace{-0.5mm} \max_{k \in \deviceset} \hspace{-0.5mm}
	\bigg[ \hspace{-0.7mm}
	\frac{\datasize \cpunumberserver}{\sfreq} - \frac{\datasize \cpunumber}{\dfreq}	 \hspace{-0.7mm}
	\bigg], \label{server_preference}
\end{align} 
where the first term is the energy consumption for training $\sweights$ and the second term is related to the elapsed time during one global epoch. Hence, depending on the balancing parameter $\gamma$, the optimal fraction $\hat{\fraction}_S$ will be either zero or one. From \eqref{dfraction} and \eqref{server_preference}, we can obtain $\ideal$. Then, for a given $\KSBS$, we can formulate the feasibility problem as follows 
\begin{align}
&\text{Find} \quad \fraction \label{feasibility}\\
&\text{s.t.} \quad \KSBS \ideal  = (U_{d,1}(\fraction), \dots, U_{d, N}(\fraction), \sutil ).
\end{align}
Since $\dutil$ and $\sutil$ are a concave and a linear function with respect to $\fraction$, respectively, it is straightforward to find $\fraction$ such that $\dutil = \KSBS U_{d, k}^{\text{ideal}}, \forall k$ and $\sutil = \KSBS U_S^{\text{ideal}}$ using a software solver.

\begin{figure}
	\centering
	\includegraphics[width=0.65\columnwidth]{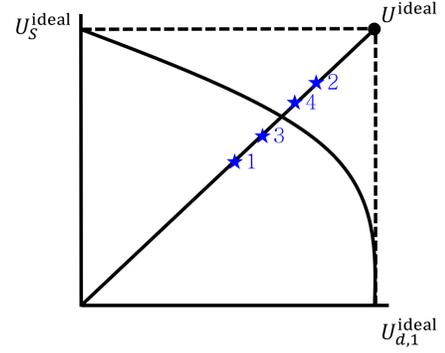}
	\captionsetup {singlelinecheck = false}

	\caption{An illustration of the Algorithm 2 for the two player case.}
	\label{fig:bisection}

	\vspace{-0.3cm}	
\end{figure}

We now obtain the \ac{KSBS} by using the bisection method with the feasibility problem \eqref{feasibility} as shown in Fig. \ref{fig:bisection} \cite{No:09}. We first set $\KSBS_{\text{max}} = 1$, $\KSBS_{\text{min}} = 0$, and $\KSBS = \frac{\KSBS_{\text{min}} + \KSBS_{\text{max}}}{2}$. Then, at iteration $n$, we solve the feasibility problem \eqref{feasibility} for $\KSBS(n)$. If it is feasible, we set $\KSBS_{\text{min}} = \KSBS(n)$. Otherwise, we set $\KSBS_{\text{max}} = \KSBS(n)$. We repeat this iteration until a certain stopping criteria becomes satisfied. The summary of our approach is provided in Algorithm \ref{alg:2}. The key complexity of Algorithm \ref{alg:2} stems from solving the feasibility problem \eqref{feasibility}. Since we should solve $\Whole$ equations in \eqref{feasibility}, the complexity of Algorithm \ref{alg:2} will be proportional to the total number of devices $\Whole$. 

In practice, we can assume that the devices send their channel information, hardware information, size of dataset, and preference toward privacy to the server through the designated interface. Then, the server can perform Algorithm \ref{alg:2}.

\begin{algorithm}[t!]
	\scriptsize
	\caption{Algorithm for the \ac{KSBS}}\label{alg:2}
	{
		Set $\KSBS_{\text{min}} = 0$ and $\KSBS_{\text{max}} = 1$ \;
		\While{$|\KSBS_{\text{max}} - \KSBS_{\text{min}}| < \epsilon$}{
			$\KSBS \leftarrow \frac{\KSBS_{\text{max}} + \KSBS_{\text{min}}}{2}$\;
			Solve the feasibility problem \eqref{feasibility} \;
			\If{$\KSBS$ is feasible}{$\KSBS_{\text{min}} \leftarrow \KSBS$ \;}
			\Else{$\KSBS_{\text{max}} \leftarrow \KSBS$ \;}
			
			}
		}
	\vspace{-0.2cm}
\end{algorithm}

\section{Simulation Results} \label{sec:simlulation results}
\begin{figure}
		\vspace{-0.35 cm}	
	\centering
	\includegraphics[width=0.75\columnwidth]{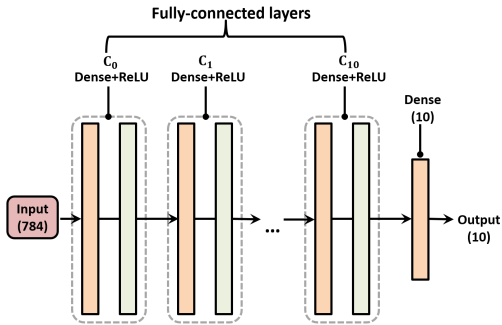}
	\captionsetup {singlelinecheck = false}

	\caption{\ac{MLP} model architecture with one input layer, 11 fully connected layers, and one output layer.}
	\label{fig:MLP}

	\vspace{-0.1 cm}	
\end{figure}

For our simulations, we distribute $\Whole = 10$ devices uniformly over a $50$ m $\times$ $50$ m square area and locate the server at the center. We adopt a Rayleigh fading channel model with a path loss exponent of 4 between the devices and the server. For a default setting, we use $\txpower = 100$ mW, $\bandwidth = 10$ MHz, $\Noise = -174$ dBm, and $\cpuparam = 2\times10^{-28}$. $\dfreq$ follows uniform distribution between $(1.5, 2.4)$ GHz, $\privacy$ is uniformly distributed between $(25, 30)$, and $\payoff$ follows uniform distribution between $(10^{-8}, 10^{-7})$. We also set $\cpunumber = \cpunumberserver = 10^3$, $\forall k$, $\budget = 1215$, $\gamma = 0.01$, and $\sfreq = 4$ GHz \cite{Ngutra:19} \cite{Zhaya:21}. We use \ac{MLP} model to classify 10 digits and clothes in the MNIST and FMNIST datasets, respectively. The model consists of one input layer, 11 fully-connected layers blocks, $C_0, C_1, \dots, C_{10}$, and one classification layer as shown in Fig. \ref{fig:MLP}. Each block $C_k$ consists of one dense layer and ReLU activation. The total number of model parameters is $|\weights| = 287955$. We split both the MNIST/FMNIST dataset into $55000$ samples for training, $5000$ samples for validation, and $10000$ samples for testing. We distribute the training dataset over devices in non-iid fashion. We choose two major and eight minor labels for each device. Then, we allocate $40\%$ of each major label and $5\%$ of each minor label to a device. We also distribute the validation/test datasets over devices using the same method as the training dataset \cite{Fa:20}. We use Adam optimizer with learning rate $0.01$ and mini-batch size is 256. For each global round, each device runs $\SGDrun = 25$ local steps.

From the given setting, our \ac{KSBS} is $\fraction^* = 0.379$ and this corresponds to $C_3$, which becomes the cut layer. Hence, the input layer up to the cut layer $C_3$ will be assigned to the device-side model $\dweights, \forall k$ with $|\dweights| = 117135$, and all subsequent layers are assigned to the server-side model $\sweights$ with $|\sweights| = 170820$. All statistical results are averaged over a large number of independent runs.

\begin{table}[t!] 
	\larger
	\centering
	\begin{tabular}{|c|c|c|}
		\hline
		Algorithms & MNIST   & FMNIST \\ \hline
		Proposed   & 93.52\% & 92.01\%        \\ \hline
		SplitFed   & 92.90\% & 79.65\%        \\ \hline
	\end{tabular}
	\caption{Performance of different algorithms on test dataset}
	\label{tab:acc}
	\vspace{-0.3cm}
\end{table}

To benchmark our proposed learning algorithm, we use SplitFed \cite{SplitFed} as a baseline. In SplitFed, FedAvg is performed on both device-side models and server-side models for every global round while our proposed algorithm only averages the server-side models. Specifically, after the server performs FedAvg on $\weights_{S,k}(t), \forall k$, each device $k$ transmits its device-side model $\dweights(t)$ to an edge server for averaging. Note that the edge server only does FedAvg on $\dweights(t)$ and does not perform forward/back propagation. Subsequently, the Fed server generates $\weights_d(t+1) = \frac{1}{\Whole}\sum_{k\in\deviceset}{\dweights(t)}$ and broadcasts it to devices. Then, devices set $\dweights(t+1) = \weights_d(t+1)$ for the next global round.

\begin{figure}[t!]
	\centering
	\includegraphics[width=0.80\columnwidth]{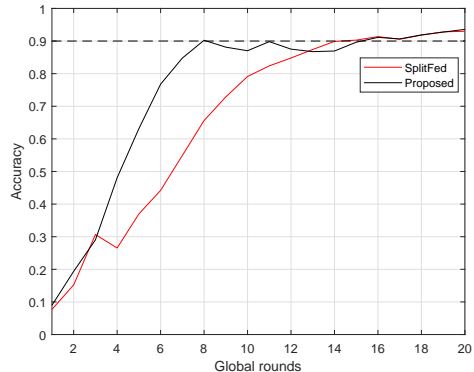}
	\captionsetup {singlelinecheck = false}

	\caption{Validation accuracy of the proposed algorithm and SplitFed on non-iid MNIST dataset}
	\label{fig:MNIST}

	\vspace{-0.3 cm}	
\end{figure}

\begin{figure}[t!]
	\centering
	\includegraphics[width=.80\columnwidth]{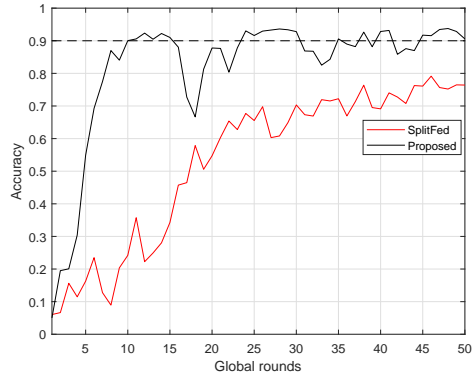}
	\captionsetup {singlelinecheck = false}

	\caption{Validation accuracy of the proposed algorithm and SplitFed on non-iid FMNIST dataset}
	\label{fig:FMNIST}

	\vspace{-0.3 cm}	
\end{figure}

Figures \ref{fig:MNIST} and \ref{fig:FMNIST} show the accuracy on the MNIST/FMNIST validation datasets as a function of global rounds for our algorithm and SplitFed. In Figs. \ref{fig:MNIST} and \ref{fig:FMNIST}, we can see that the proposed algorithm converges faster than the baseline on both datasets. From Table \ref{tab:acc}, we observe that, although the baseline achieves similar performance with the proposed algorithm on the MNIST test dataset, it does not perform well on more difficult dataset, which is FMNIST. Meanwhile, our algorithm shows more robust accuracy on both non-iid datasets. This is because the proposed algorithm can mitigate discrepancies among the individual device optimum via personalization. Unlike the baseline, our algorithm only averages the server-side models while keeping the device-side models personalized. Then, each device-side model can move toward its local optimum during training. Therefore, it can achieve fast convergence as well as generalization through the server-side models. Meanwhile, SplitFed averages all layers and then moves toward the average of all individual optimum points resulting in slow convergence \cite{scaffold}.

\begin{figure}[t!]
	\centering
	\begin{subfigure}[t]{0.40\textwidth}
		\centering
		\includegraphics[width=\textwidth]{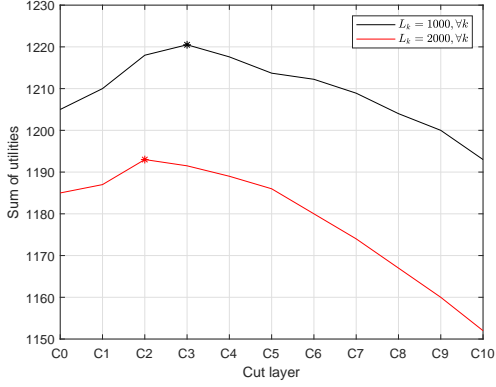}
		\caption{$\boldsymbol{\lambda} \sim U(25, 30)$}
		\label{fig:lower_priacy}
		\vspace{-0.cm}
	\end{subfigure}
	\vspace{-0.cm}
	\begin{subfigure}[t]{0.40\textwidth}
	    \centering
		\includegraphics[width=\textwidth]{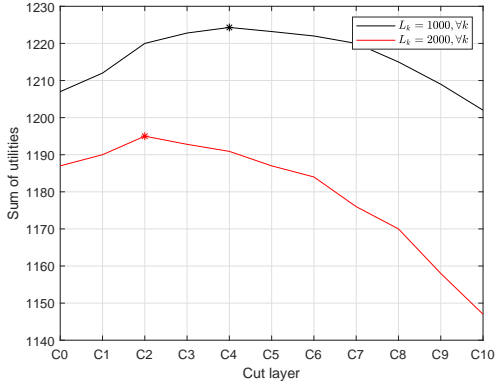}
		\caption{$\boldsymbol{\lambda} \sim U(30, 35)$}
		\label{fig:higher_priacy}
	\end{subfigure}
	\caption{Sum of utilities with different privacy parameter distributions}
	\label{fig:privacy}
	\vspace{-0.3cm}
\end{figure}

Figure \ref{fig:privacy} presents the sum of utilities for each cut layer with different distribution of privacy parameters $\boldsymbol{\lambda} = \{\lambda_1, \dots, \lambda_\Whole\}$. From Fig. \ref{fig:lower_priacy}, we can clearly see that our cut layer $C_3$, which is obtained from the \ac{KSBS}, can achieve the best sum of utilities. Moreover, as the number of required CPU cycles to process one data sample $\cpunumber$ increases, we can see that the optimal cut layer decreases. This is because devices have to spend more energy for training, so having a large device-side model is not beneficial. This also corroborates \eqref{dfraction}, which shows that the optimal cut layer for each device is a decreasing function of $\cpunumber$. In Fig. \ref{fig:higher_priacy}, $\privacy, \forall k$ follows uniform distribution between $[30, 35]$ resulting in a stronger privacy preference for all devices than Fig. \ref{fig:lower_priacy}. From the given setting, the \ac{KSBS} is found to be $0.506$, and this corresponds to $C_4$ for the cut layer. We can see that the optimal cut layer increased to $C_4$ from $C_3$. This is because devices now have a stronger preference for data protection and have more bargaining power due to the monotonicity axiom of the \ac{KSBS}. 

\section{Conclusion} \label{sec:conclusion}
\vspace{-0.0cm}
In this paper, we have studied the problem of finding the optimal split on a neural network in a personalized \ac{SL} over wireless networks. We have presented the training algorithm for the proposed personalized \ac{SL} to tackle non-iid datasets. We also have introduced utility functions by considering energy consumption, training time, and data privacy during training. Then, we have formulated a multiplayer bargaining problem to find the optimal cut layer between devices and the server to maximize their utilities. To solve the problem, we have obtained the \ac{KSBS} using the bisection method and the feasibility test. Our simulation results have shown that the proposed learning algorithm can converge faster than the baseline and the \ac{KSBS} can provide the best sum utilities. Moreover, we have shown that the proposed algorithm can achieve significantly higher accuracy in non-iid datasets. 
\vspace{-0.0cm}

\bibliographystyle{IEEEtran}
\bibliography{Bibtex/StringDefinitions,Bibtex/IEEEabrv,Bibtex/mybib}

\end{document}

%% file: SupportDocuments/acronym.tex
\acrodef{IoT}{Internet of Things}
\acrodef{BS}{base station}
\acrodef{pdf}{probability density function}
\acrodef{i.i.d.}{independent and identically distributed}
\acrodef{CDF}{cumulative distribution function}
\acrodef{FL}{federated learning}
\acrodef{ML}{machine learning}
\acrodef{SGD}{stochastic gradient descent}
\acrodef{MAC}{multiply-accumulate}
\acrodef{CNN}{convolutional neural network}
\acrodef{DNN}{deep neural network}
\acrodef{SL}{split learning}
\acrodef{OFDMA}{orthogonal frequency domain multiple access}
\acrodef{NBS}{Nash bargaining solution}
\acrodef{KSBS}{Kalai-Smorodinsky bargaining solution}
\acrodef{MLP}{multi-layer perceptron}

%% file: SupportDocuments/defmetric.tex
\usepackage{color}
\usepackage{dsfont}
\usepackage{bbm}

\newtheorem{theorem}{Theorem}
\newtheorem{lemma}{Lemma}

\newcommand{\E}{\mathbb{E}}

\newcommand{\Whole}{N}
\newcommand{\weights}{\boldsymbol{w}}

\newcommand{\datasize}{D_k}
\newcommand{\Wholedata}{D}
\newcommand{\dataset}{\mathcal{D}_k}
\newcommand{\loss}{\ell}

\newcommand{\SGDrun}{I}
\newcommand{\ka}{\nonumber \\}

\newcommand{\minibatch}{\xi_k}
\newcommand{\dweights}{\weights_{d, k}}
\newcommand{\alldweights}{\weights_d}
\newcommand{\sweights}{\weights_S}
\newcommand{\sdweights}{\weights_{S,k}}
\newcommand{\fraction}{\alpha}
\newcommand{\activation}{a_{d,k}}
\newcommand{\dlabels}{Y_k}
\newcommand{\deviceset}{\mathcal{N}}
\newcommand{\payoff}{c_k}
\newcommand{\dfreq}{f_k}
\newcommand{\dutil}{U_{d,k}(\fraction)}
\newcommand{\sutil}{U_{S}(\fraction) }
\newcommand{\cpuparam}{\kappa}
\newcommand{\privacy}{\lambda_k}
\newcommand{\cpunumber}{L_k}
\newcommand{\cpunumberserver}{L_S}
\newcommand{\budget}{B}
\newcommand{\sfreq}{f_S}
\newcommand{\balance}{\gamma}
\newcommand{\denergy}{E^C_k(\fraction)}
\newcommand{\tenergy}{E^U_k}
\newcommand{\senergy}{E_S(\fraction)}
\newcommand{\stime}{T_S(\fraction)}
\newcommand{\dtime}{T_k(\fraction)}
\newcommand{\ttime}{\tau_k}
\newcommand{\achievable}{R_k}
\newcommand{\bandwidth}{W}
\newcommand{\txpower}{P_k}
\newcommand{\channel}{h_k}
\newcommand{\Noise}{N_0}
\newcommand{\utilset}{\mathcal{U}}
\newcommand{\disagreementset}{\boldsymbol{\phi} } 
\newcommand{\disagreement}{\boldsymbol{\phi}}
\newcommand{\barginingsolution}{f}
\newcommand{\ideal}{\boldsymbol{U}^{\text{ideal} } } 
\newcommand{\KSBS}{\beta}